\documentclass[english]{article}
\usepackage[T1]{fontenc}
\usepackage[latin1]{inputenc}
\usepackage{graphicx}
\usepackage{amssymb}

\makeatletter

\newcommand{\lyxaddress}[1]{
\par {\raggedright #1
\vspace{1.4em}
\noindent\par}
}

\usepackage{babel}
\makeatother
\begin{document}

\title{Black Hole evaporation in a thermalized final-state projection
model}

\author{A. Fabbri and A. Pérez}

\maketitle

\lyxaddress{\begin{center}Departamento de F\'{\i}sica Teórica and IFIC, \\
Universidad de Valencia-CSIC, \\
Dr. Moliner 50, 46100 Burjassot, Spain\par\end{center}}

\begin{abstract}
We propose a modified version of the Horowitz-Maldacena
final-state boundary condition based upon a matter-radiation
thermalization hypothesis on the Black Hole interior, which
translates into a particular entangled state with thermal Schmidt
coefficients. We investigate the consequences of this proposal for
matter entering the horizon, as described by a Canonical density
matrix characterized by the matter temperature $T$. The emitted
radiation is explicitly calculated and is shown to follow a
thermal spectrum with an effective temperature $T_{eff}$. We
analyse the evaporation process in the quasi-static approximation,
highlighting important differences in the late stages with respect
to the usual semiclassical evolution, and calculate the fidelity
of the emitted Hawking radiation relative to the infalling matter.
% indicates that information
%entering the horizon would be released when the Hawking
%temperature of the evaporating black hole becomes a fraction
%($\sim10\%$ ) of the collapsing matter temperature. We analyze the
%implications of this result for a typical astrophysical scenario.
\end{abstract}
Black Holes (BHs) are probably the most fascinating objects in our
Universe. Although initially related to the General Theory of
Relativity, they have become an interdisciplinary field, where
ideas from Thermodynamics, Quantum Field Theory in Curved
Spacetimes, String Theory and Quantum Information are applied in
order to understand those aspects which go beyond the purely
gravitational context. Among them, the most spectacular is the BH
evaporation effect \cite{Hrad}, which shows that quantum BHs,
unlike their classical counterparts, emit particles in the form of
thermal radiation. Even more intriguing is the suggestion Hawking
made \cite{Hawk76} that black holes will evaporate completely and
the information about their formation will be lost forever.
Obviously, if the whole process is governed by a unitary
transformation, as demanded by Quantum Mechanics, no information
can be lost. This question, however, is still under debate, and
several hypothesis have been suggested \cite{muchos}.

Recently, Horowitz and Maldacena \cite{HM} (HM, hereafter) have made
a proposal to describe this transformation based on a final-state
projection condition, which resembles quantum teleportation (but without
a classical communication channel). In the original HM proposal, matter
inside the BH is in a maximally entangled state with the incoming
Hawking radiation. The unknown effects of quantum gravity are encoded
into an additional unitary transformation $S$ acting on matter states
alone. The overall transformation acting on matter entering the horizon
is obtained by the projection onto this final state.

Several comments and modifications of this interesting proposal
have been discussed later. Gottesmann and Preskill \cite{GP} have
argued that the interaction between the collapsing body and the
infalling Hawking radiation would modify the suggested maximally
entangled state, and this modification gives rise, in general, to
a loss of information. Following this idea, in \cite{Lloyd} the
author allows for a modified final state with random Schmidt
coefficients. As a consequence, some information is lost, giving
{}``almost certain escape'' for general states.

In this letter, we propose a different form for the HM entangled
state, which is in fact described by a thermal spectrum.  Several
arguments will be given to justify this proposal. We will then
determine explicitly the radiation emitted by the BH, and
analyze both the consequences regarding the evaporation process
and the relation between the information contained in the infalling matter
and that of the emitted Hawking radiation.
%A discussion concerning the
%interpretation of the results obtained will be made at the end.
We use units such that $G=\hbar=c=k_{B}=1$.

Let us begin with a brief description of the final-state
projection proposal. The Hilbert space of the infalling matter
plus Hawking radiation can be written as

\begin{equation}
H=H_{M}\otimes H_{in}\otimes H_{out}\end{equation} where $H_{M}$ corresponds
to matter degrees of freedom, and $H_{in}$ ($H_{out}$) stands for
the incoming (outgoing) Hawking radiation, described by an
entangled thermal state belonging to $H_{in}\otimes H_{out}$. We
denote this state by $|\phi\rangle_{in\otimes out}$ and can be
expressed, in a compact form, as

\begin{equation}
|\phi\rangle_{in\otimes
out}=\sum_{j}\lambda_{j}^{T_{H}}|j\rangle_{in}\otimes|j\rangle_{out}.\label{inout}\end{equation}

In the above equation, the states $|j\rangle_{in}$
($|j\rangle_{out}$) denote Fock states for the incoming (outgoing)
radiation. In a more detailed way, we can write

\begin{equation}
|j\rangle_{in}=|N_{1}N_{2}...\rangle_{in}\label{Fockstate}\end{equation}
 (analogously for $|j\rangle_{out}$ ), where $N_{i}$ denotes the occupation
number for frequency $\omega_{i}$ , ($i=1,2,...$ ) and we have to
take into account all possible positive frequencies. Following these
conventions, the coefficients in Eq. (\ref{inout}) would read as
\begin{equation}
\lambda_{j}^{T_{H}}=\frac{1}{\sqrt{Z_{H}}}\exp(-\beta_{H}\epsilon_{j}/2),\label{lambdaT}\end{equation}
with $\epsilon_{j}$ the energy corresponding to the Fock state 
(\ref{Fockstate}), $\beta_{H}=1/T_{H}$ the inverse of the Hawking
temperature, given, for the Schwarzschild BH, by
$T_{H}=1/8\pi M$ (M is its mass), and
$\frac{1}{\sqrt{Z_{H}}}$ is a normalization factor.

Following HM, the matter state inside the BH, together with the
infalling radiation, form a pure entangled state
$|\psi\rangle_{M\otimes in}$, written in the form

\begin{equation}
|\psi\rangle_{M\otimes
in}=\sum_{j}\lambda_{j}|j\rangle_{M}\otimes|j\rangle_{in}.\label{psiMin}\end{equation}

The states $\{|j\rangle_{M}\}$ form an orthonormal basis of the
Hilbert space $H_{M}$ . In the original proposal, it is suggested
that this state is maximally entangled, therefore all the
$\lambda_{j}$ would be the same. Moreover, as mentioned above, a
BH unitary transformation $S$ acting on matter degrees of freedom
is included in that proposal. We will, however, keep the above
form, with yet undefined coefficients $\lambda_{j}$, as a starting
point for our discussion.

Let us first compare Eqs. (\ref{inout}) and (\ref{psiMin}). From
each of them one can obtain the density matrix describing the infalling
Hawking radiation, by tracing out over the remaining degrees of freedom.
For consistency, both expressions should coincide, i.e.:

\begin{equation}\rho_{in}=tr_{out}(|\phi\rangle_{in\otimes
out}\,\,\,_{in\otimes
out}\langle\phi|)=tr_{M}(|\psi\rangle_{M\otimes
in}\,\,\,_{M\otimes in}\langle\psi|).\end{equation}

In other words,

\begin{equation}
\rho_{in}=\sum_{j}(\lambda_{j}^{T_{H}})^{2}|j\rangle_{in}\,\,_{in}\langle
j| =\sum_{j}(\lambda_{j})^{2}|j\rangle_{in}\,\,_{in}\langle j| \end{equation}
which implies $\lambda_{j}=\lambda_{j}^{T_{H}}$ $\forall j$ . As a
consequence, the state $|\psi\rangle_{M\otimes in}$ described by
Eq. (\ref{psiMin}) adopts the form of a thermal state, similar to
Eq. (\ref{inout}). More precisely, if we trace out the incoming
radiation, the resulting density matrix is given by

\begin{equation}
\rho_{M}=tr_{in}(|\psi\rangle_{M\otimes in}\,\,\,_{M\otimes
in}\langle\psi|)=\sum_{j}(\lambda_{j}^{T_{H}})^{2}|j\rangle_{M}\,\,_{M}\langle
j|,\end{equation} where the diagonal elements $(\lambda_{j}^{T_{H}})^{2}$
follow a (Canonical) thermal distribution with temperature equal
to the Hawking temperature. Let us discuss this result in more
detail. One could argue that, as matter enters the horizon, it
would eventually thermalize with the incoming radiation, and
finally adopt a thermal distribution with the same temperature
(the Hawking temperature). A supporting argument for this
hypothesis is given in \cite{BF06}, where it is claimed that, as a
space-like singularity is approached, solutions to Einstein's
equations become chaotic, with rapid cycles through all states in
the Hilbert space, so that time averages give the same results as
ensemble averages on a thermal system.

We also note that the above suggested (thermal) form for
$|\psi\rangle_{M\otimes in}$ has also been discussed in \cite{Ahn}
within the context of gravitational collapse of a matter shell
(described by a scalar massless field). It is then shown that, for
a given frequency $\omega$, the stationary state of matter and
incoming radiation inside the black hole is a {}``maximally
entangled two-mode squeezed'' state with coefficients given by Eq.
(\ref{lambdaT}). However, when discussing the final state
projection, in this reference the author also includes the unknown
$S$ matrix discussed in the HM proposal. Our point of view is
different: given the arguments above, we suggest the state
$|\psi\rangle_{M\otimes in}$ as the state to be used for the final
state projection. In this way, our proposal resembles the one in
\cite{Lloyd} (compare Eq. (6) in this reference with our Eq.
(\ref{psiMin})), but now the $\lambda_{j}$'s to be used, instead
of possessing a random distribution, have a thermal one, shown in
(\ref{lambdaT}).

This specific form for the final state allows us to make explicit
calculations for the emitted radiation within the context of final
state projection, as suggested by HM. Following these ideas, we define
the projector

\begin{equation}
\Lambda_{M\otimes in}=|\psi\rangle_{M\otimes in}\,\,\,_{M\otimes
in}\langle\psi|.\end{equation}

An incoming state of matter $|\chi\rangle_{M}$ entering the
horizon would be transformed into the state $|\phi\rangle_{out}$
of outgoing Hawking radiation, according to the projection

\begin{equation}
\Lambda_{M\otimes in}(|\chi\rangle_{M}|\phi\rangle_{in\otimes
out})=|\psi\rangle_{M\otimes in}|\phi\rangle_{out}.\end{equation}

In order to study more realistic situations, we generalize the
above rule to matter described by a density matrix. This would be
the case, for example, when we consider a nondegenerate gas
falling into the black hole, as we discuss later. Let $\rho_{M}$
be the matter density matrix, and define $\rho_{in\otimes
out}=|\phi\rangle_{in\otimes out}\,\,\,_{in\otimes
out}\langle\phi|$. We extend the final state projection in the
straightforward way

\begin{equation}
\Lambda_{M\otimes in}(\rho_{M}\otimes\rho_{in\otimes
out})\Lambda_{M\otimes in} \equiv|\psi\rangle_{M\otimes
in}\,\,\,_{M\otimes in}\langle\psi|\otimes\bar{\rho}_{out},\end{equation}
where $\bar{\rho}_{out}\equiv{}_{M\otimes
in}\langle\psi|(\rho_{M}\otimes\rho_{in\otimes
out})|\psi\rangle_{M\otimes in}$ is the unnormalized density
matrix describing the outgoing Hawking radiation. After some
algebra, one obtains the following result for the normalized
density matrix:

\begin{equation}
\rho_{out}=\frac{\sum_{i,j}\rho_{Mij}\lambda_{i}^{T_{H}}\lambda_{j}^{T_{H}}|i\rangle_{out}\,\,\,_{out}\langle
j|}{\sum_{i}\rho_{Mii}(\lambda_{i}^{T_{H}})^{2}},\end{equation} with
$\rho_{Mij}={}_{M}\langle i|\rho_{M}|j\rangle_{M}$ the matrix
elements of $\rho_{M}$ . As a particular case, which will be
useful for further discussions, let us assume that the infalling
matter is described by a Canonical ensemble characterized by a
temperature $T$, such that
$\rho_{M}=\frac{1}{Z}\sum_{i}e^{-\beta\epsilon_{i}}|i\rangle_{M}\,\,_{M}\langle
i|$ and $\beta=1/T$ . Here $Z=\sum_{i}e^{-\beta\epsilon_{i}}$ is
the matter partition function. In this case, it is straightforward
to obtain

\[
\rho_{out}=\frac{1}{Z_{eff}}\sum_{i}e_{i}^{-\beta_{eff}\epsilon_{i}}|i\rangle_{out}\,\,\,_{out}\langle
i|,\] where $\beta_{eff}=\beta+\beta_{H}$ and
$Z_{eff}\equiv\sum_{i}e_{i}^{-\beta_{eff}\epsilon_{i}}$. From the
above equation, we immediately see that, as a consequence of the
model introduced in this paper, the BH radiates with an effective
temperature \begin{equation} \label{Teff}
T_{eff}=\frac{TT_{H}}{T+T_{H}}.\end{equation}

Suppose we consider the infall of matter inside a BH. This
situation could correspond to the accretion disk around a solar
mass (or larger) BH. The gas can be described by a non-degenerate
(Canonical) distribution function. The temperature of the gas in
the inner disk depends on several parameters such as the mass of
the BH, the distance to the center and the viscosity of the gas,
but typically one finds temperatures $T\sim 10^{7}K$
\cite{Shapiro}. For $M \sim M_{\odot}$, the Hawing temperature is
$T_{H}\sim10^{-7}K$. Since $T\gg T_H$ the black hole emits
radiation at the usual Hawking temperature  $T_{eff}\sim T_H$.
Using standard arguments, let us now consider the evaporation
process modeled  as a sequence of quasi-stationary states, each of
them radiating with the instantaneous temperature
$T_{eff}=T_{eff}(T, T_H(M(t)))$, which varies with time, and where
the BH mass $M(t)$ satisfies the evolution equation

\begin{equation}
\frac{dM}{dt}=-4\pi R_{S}^{2}\sigma T_{eff}^{4},\end{equation} with $R_{s}=2M$
the Schwarzschild radius and
%\footnote{For simplicity we consider
%only photons; it is straightforward to extend the analysis to
%include all kinds of particles.} $\sigma=\frac{\pi^{2}}{60}$
$\sigma$ is the Stefan-Boltzmann constant. In the usual case
$T_{eff}\to T_H$ and we have that the BH cooling time
(corresponding to complete evaporation) is
$t_{c}=\frac{256\pi^{3}}{3\sigma}M_{0}^{3}$, where $M_{0}=M(0)$ is
the BH initial mass. Now, defining $m=M(t)/M_{0}$ and
$\tau=t/t_{c}$ the above equation can be rewritten as

\begin{equation}
\frac{dm}{d\tau}=-\frac{\gamma^{4}m^{2}}{3(1+\gamma
m)^{4}},\label{cooling}\end{equation} where $\gamma=T/T_{H}^{0}$
and $T_{H}^{0}=1/8\pi M_0$ is the initial Hawking temperature.

Eq. (\ref{cooling}) shows two different regimes. The first one
corresponds to $\gamma m=T/T_H(t)\gg1$ . During this phase one has
$T_{eff}\simeq T_{H}$ and the evaporation proceeds as in the
standard case, i.e. $m^3 \simeq 1 - \tau$. However, when
$m\gamma\ll1$, that is $T\ll T_H(t)$ and $T_{eff}\simeq T$, the
evolution is drastically modified to $m\simeq 3/\gamma^{4}\tau$,
i.e. the mass goes to zero asymptotically, rather than showing a
finite evaporation time.

For the example described above, one has $\gamma\sim10^{14}$ ,
which means that the transition $\gamma m\sim 1$ occurs at $\tau =
\tau_1\sim 1-10^{-42}$, when the BH mass is $M\sim
M_{0}/\gamma\sim10^{19}g.$ Such a mass is certainly macroscopic,
and well above the Planck mass scale, where quantum gravity
effects are thought to dominate the evaporation. Equivalently, the
transition corresponds to $T_{H}\sim T$. As we show below, at this
stage the information released by the evaporating BH starts to
approach the one corresponding to the initial infalling matter.

In order to compare how much information have in common the
incoming matter and the outgoing radiation, we compute the
fidelity
$F(\rho_{out},\rho_{M})=tr[\rho_{M}^{1/2}\rho_{out}\rho_{M}^{1/2}]^{1/2}$
between $\rho_{M}$ and $\rho_{out}$ . Notice, however, that these
two operators are expressed in different basis. Therefore, we
first add the operator $U'$ which performs the trivial,
information conserving map $U'|i\rangle_{M}=|i\rangle_{out}$ .
After some simple algebra, one obtains

\begin{equation}
F(\rho_{out},\rho_{M})=\frac{1}{(ZZ_{eff})^{1/2}}\sum_{j}e^{-(\beta+\beta_{eff})\epsilon_{j}/2}.\label{fidelity}\end{equation}

It is simpler to take logs on the above expression. For example,
one has $\log Z=-\sum_{\omega}\log(1-e^{-\beta\omega})$. We
perform the sum over frequencies by using a simple box
normalization, i.e. $\sum_{\omega}\longrightarrow
V\int\frac{d^{3}\omega}{(2\pi)^{3}}$ , where $V$ is the box
volume, which we take as $V=\frac{4}{3}\pi R_{s}^{3}$ .
%We will come back to this point in our final discussion.
The remaining
terms in Eq. (\ref{fidelity}) are calculated in a similar way,
giving the final expression

\begin{equation}
\log
F(\rho_{out},\rho_{M})=-\frac{24y^{4}+48y^{3}+33y^{2}+9y+1}{8640y^{3}(2y^{2}+3y+1)^{3}}.
\label{fidelityfin}\end{equation} The latter formula depends only
on the ratio $y=T_{H}/T$ .

\begin{figure}
\includegraphics[width=8cm]{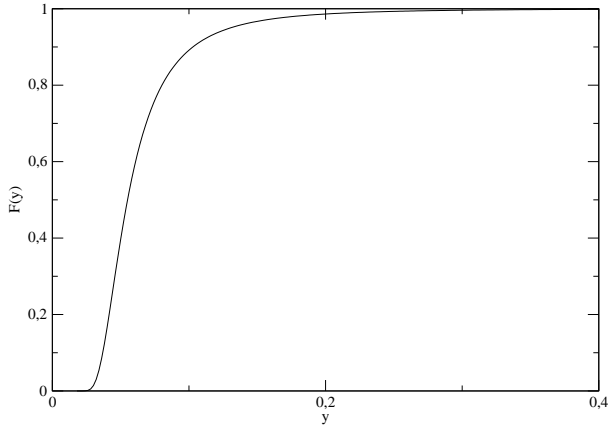}

\caption{Fidelity of $\rho_{out}$ with respect to $\rho_{M}$, as
calculated from Eq. (\ref{fidelityfin}).}
\end{figure}

The resulting expression, Eq. (\ref{fidelityfin}), is plotted in
Fig. 1, as a function of $y$ for the most relevant range. As it
follows from the above formula, $F(\rho_{out},\rho_{M})$ is
exponentially suppressed for $T\gg T_{H}$ , and becomes close to
unity as $T_{H}$ approaches $T$. Let us return to the
astrophysical scenario described previously. Initially, we have
$y=1/\gamma\sim10^{-14}$; therefore, for similar scenarios
$F(\rho_{out},\rho_{M})\longrightarrow0$ , and the Hawking
radiation does not contain the information about the infalling
matter. Now assume that the accretion eventually stops, and that
the BH continues to evaporate, so that $M$ decreases and $T_{H}$
increases accordingly. Following the above results, the
information carried out by the accreted mass will be approximately
recovered when $T\sim T_{H}$ or, in other words, when the BH
approaches the asymptotic evaporation phase.

% \textit{Discussion and conclusions}

To sum up the results presented in this letter, using a
thermalized final-state projection model, slightly different from
the original proposal by Horowitz and Maldacena, we considered the
infall of matter (in the form of a thermal gas characterized by a
temperature $T$) into the Black Hole. Our model allows to
quantitatively determine the form of the radiation emitted by the
Black Hole, which is thermal at the temperature $T_{eff}$ given in
eq. (\ref{Teff}). Modeling the evaporation process as a sequence
of quasi-stationary states, each characterized by the
instantaneous temperature $T_{eff}$, we have shown that, for
realistic values of the Black Hole initial mass $M_0$ and of $T$,
the emission rate is the standard one, i.e. $T_{eff}\sim T_H$, for
most of its lifetime.

Extrapolation of our results to the late stages of the evaporation
shows the transition to a new regime, where $T_{eff}$ reaches its
maximum value $T$ and the black hole mass goes to zero
asymptotically. The fidelity, measuring the information content in
the emitted radiation relative to the initial infalling matter, is
essentially zero during the first phase of the evaporation, and
goes to unity as the BH approaches the asymptotic regime, when
information from the infalling matter would be released.

\textbf{Acknowledgments}

This work has been supported by the Spanish Grants AYA2004-08067-C01,
FPA2005-00711, FIS2005-05736-C03-03, the EU Network MRTN-CT-2004-005104
and by Generalitat Valenciana (Grant GV05/264).


\begin{thebibliography}{1}
\bibitem{Hrad}S.W. Hawking, \textit{Comm. Math. Phys.} \textbf{43}
(1975) 199.

\bibitem{Hawk76}S.W. Hawking, \textit{Phys. Rev.} \textbf{D14} (1976),
2460.

\bibitem{muchos}See for instance: D.N. Page, \textit{Phys. Rev.
Lett. } \textbf{44} (1980), 301; S.B. Giddings, Quantum mechanics
of black holes, {[}hep-th/9412138]; T. Banks, A. Dabholkar, M.R.
Douglas and M. O'Loughlin, \textit{Phys. Rev. } \textbf{D45}
(1992), 3607; S.W. Hawking, \textit{Phys. Rev. } \textbf{D37}
(1988), 904; A. Strominger, Les Houches lectures on black holes,
{[}hep-th/9501071]; J.D. Bekenstein, Black hole hair: 25 years
after, {[}gr-qc/9605059]; 't Hooft, \textit{Nucl. Phys. }
\textbf{B335} (1990), 138; L. Susskind, L. Thorlacius and J.
Uglum, \textit{Phys. Rev. } \textbf{D48} (1993), 3743

\bibitem{HM}G. T. Horowitz and J. Maldacena, \textit{J. High Energy
Phys.} \textbf{02} (2004) 008 {[}hep-th/0310281].

\bibitem{GP}D. Gottesman and J. Preskill, \textit{J. High Energy
Phys.} \textbf{03} (2004) 026 {[}hep-th/0311269].

\bibitem{Lloyd}S. Lloyd, \textit{Phys. Rev. Lett.} \textbf{96} (2006)
061302.

\bibitem{BF06}T. Banks and W. Fischler, Space-like Singularities
and Thermalization, {[}hep-th/0606260].

\bibitem{Ahn}D. Ahn, On the final state boundary condition of the
Schwarzschild black hole {[}hep-th/0606028].

\bibitem{Shapiro}S.L. Shapiro and S.A. Teukolsky, \textit{Black Holes,
White Dwarfs, and Neutron Stars}, John Wiley and Sons (1983).
\end{thebibliography}
\end{document}